\documentclass[a4paper,12pt]{article}
\usepackage{amsmath}
\usepackage{amssymb}
\usepackage{latexsym}
\usepackage{lscape}
\usepackage{pdflscape}
\usepackage{hyperref}
\newcommand{\be}{\begin{eqnarray}}
\newcommand{\ee}{\end{eqnarray}}
\newcommand{\bea}{\begin{aligned}}
\newcommand{\eea}{\end{aligned}}
\newcommand{\bs}{\boldsymbol}
\newcommand{\ph}{\phantom}
\newcommand{\bsub}{\begin{subequations}}
\newcommand{\esub}{\end{subequations}}
\newcommand{\non}{\nonumber}
\newcommand{\ket}[1]{|#1\rangle}
\newcommand{\bra}[1]{\langle #1|}

\newcommand{\disfrac}[1][2]{\displaystyle\frac}
\newtheorem{Theo}{Theorem}[section]
\begin{document}
\title{\large\textbf{Lie algebra automorphisms as Lie point symmetries and the solution space for Bianchi Type $I, II, IV, V$ vacuum geometries}}
\author{\textbf{Petros A. Terzis}\thanks{pterzis@phys.uoa.gr} ~\textbf{and} \textbf{T. Christodoulakis}\thanks{tchris@phys.uoa.gr}\\
University of Athens, Physics Department\\
Nuclear \& Particle Physics Section\\
Panepistimioupolis, Ilisia GR 157--71, Athens, Hellas}
\date{}
\maketitle
\begin{center}
\textit{}
\end{center}
\vspace{0.5cm} \numberwithin{equation}{section}
%%%%%%%%%%%%%%%%%%%%%%%%%%%%%%%%%%%%%%%%%%%%%%%%%%%%%%%%%%%%%%%%%%%%%%%%%%%%%%%%%%%%%%%%
\begin{abstract}
Lie group symmetry analysis for systems of coupled, nonlinear ordinary differential equations is performed in order to obtain the entire solution
space to Einstein's field equations for vacuum Bianchi spacetime geometries. The symmetries used are the automorphisms of the Lie algebra
of the corresponding three-dimensional isometry group acting on the hyper-surfaces of simultaneity for each Bianchi Type, as well as the scaling
and the time reparametrization symmetry. A detailed application of the method is presented for Bianchi Type $IV$. The result is the acquisition of the general solution of Type $IV$ in terms of sixth Painlev\'{e} transcendent $\mathbf{P_{VI}}$, along with the known pp-wave solution. For Bianchi Types $I, II, V$ the known entire solution space is attained and very briefly listed, along with two new Type $V$ solutions of \emph{Euclidean} and \emph{neutral signature} and a Type $I$ \emph{pp-wave} metric.

\end{abstract}
%%%%%%%%%%%%%%%%%%%%%%%%%%%%%%%%%%%%%%%%%%%%%%%%%%%%%%%%%%%%%%%%%%%%%%%%%%%%%%%%%%%%%%%%
\newpage
\section{Introduction}
The exploitation of the group of automorphisms in order to obtain a
unified treatment of spatially homogeneous Bianchi spacetime
geometries has a rather long history as it goes back to the early
1960's \cite{Heckmann}. In 1979, Harvey \cite{Harvey} found the
automorphisms of all three-dimensional Lie algebras, while the
corresponding results for the four-dimensional Lie algebras have
been presented in \cite{tchris1}. In Jantzen's tangent space
approach the automorphism matrices are considered as the means for
achieving a convenient parametrization of a full scale factor matrix
in terms of a desired, diagonal matrix
\cite{Jantzen1,Jantzen2,Uggla}. Siklos used these time-dependent
automorphisms as a tool for the proper choice of variables aiming at
a simplification of the ensuing equations \cite{Sikl1}, while Samuel
and Ashtekar were the first to look upon automorphisms from a space
viewpoint \cite{Samuel}. The notion of \textit{Time-Dependent
Automorphism Inducing Diffeomorphisms}, i.e., coordinate
transformations mixing space and time in the new spatial coordinates
and inducing automorphic motions on the scale-factor matrix, the
lapse, and the shift vector has been developed in \cite{tchris2}.
The use of these covariances enables one to set the shift vector to
zero without destroying manifest spatial homogeneity. At this stage
one can use the rigid automorphisms, i.e., the remaining gauge
symmetry, as Lie point symmetries of Einstein's field equations in
order to reduce the order of these equations and ultimately
completely integrate them \cite{tchris3}.

The methodology applied in the present work is the Lie group symmetry analysis of differential equations. For a detailed account of this approach see, e.g., \cite{Stephani}, \cite{Olver}, \cite{Bluman}. Just to put it simply, a symmetry of a system of differential equations is a transformation mapping of any solution of the system to another solution. In other words, the symmetry group of the system is the group transforming solutions of the system to other solutions. In fact, it is the largest local group of transformations acting on the system variables. Such groups are Lie groups depending on continuous parameters and consisting of either point transformations, acting on the systems space of independent and dependent variables or, more generally, contact transformations that act also on all first derivatives of the dependent variables. The theory can be also generalized to the so called higher-order (\textit{Lie-Backlund}) symmetries, by including in the transformation of the independent variables derivatives of the dependent. In this work we use all these three kinds. In general, the aforementioned transformations are nonlinear and the main benefit of the Lie group symmetry analysis comes from the replacement of the nonlinear symmetry conditions of the system investigated by ``linear'' conditions associated with the infinitesimal generators of the symmetries. To this purpose, it is necessary first to determine a class of general admissible variable transformations and then to search for special members of this class under which the system of differential equations remains invariant. It is rather evident, that the degree of generality of the admissible transformations is proportional to the number of the existing symmetries. One should further stress, that one of the main ingredients of the method lies in the notion of ``prolongation'' of a group action on the space of derivatives of the systems' dependent variables up to any finite order. One is thus able to deal with differential equations of any order. In closing this short introduction, we must point out that the knowledge of a symmetry group of a higher order differential equation has much the same consequences as the knowledge of a symmetry group of the corresponding system of first order differential equations. In fact, we will make use of the following two important theorems (for a proof, see \cite{Olver}):

\begin{Theo}
Let
\begin{equation}
\frac{dy^\nu}{dx}=F^\nu(x,y),\quad \nu=1,\ldots,\,q
\end{equation}
be a first order system of q differential equations and suppose that G is a one-parameter symmetry group of the system. Then, there exists a change of variables $(t,u) = \psi (x,y)$ under which the system can be written as
\begin{equation}
\frac{du^\nu}{dt}=H^\nu\left(t,\,u^1,\ldots,\,u^{q-1}\right),\quad\nu=1,\ldots,\,q
\end{equation}
so that the original system is reduced to a new system of $q-1$ differential equations for $u^1, u^2,\dots, u^{q-1}$ together with the integral
\begin{equation}
u^q(t)=\int H^q\left(t,\,u^1(t),\ldots,\,u^{q-1}(t)\right)dt+c.
\end{equation}
\end{Theo}

\begin{Theo}
If the system of $q$ first order differential equations considered in {\bf Theorem 1.1} admits an $r$-parameter solvable symmetry group, then the solution of the system can be found by quadrature from the solution of a reduced system of $q-r$ first order differential equations. If the original system is invariant under a $q$-parameter solvable symmetry group, then its general solution can be found by quadrature alone.
\end{Theo}

The paper is organized as follows: in Section 2, we give our method. In Section 3, the detailed application of the method to Bianchi Type IV is presented. In Section 4, the results of the method's application to Bianchi Types I, II, and V are briefly listed along with three new solutions. Finally, some discussion and concluding remarks are given in Section 5.

\section{The Method}
It is known, that  the line element for spatially
homogeneous spacetime geometries with a simply transitive action of
the corresponding isometry group \cite{Stephani1}, \cite{Ellis},
assumes the form
\begin{equation}\label{line element}
ds^2=\left(N^\alpha N_\alpha-N^2\right)dt^2+2N_\alpha
\sigma^\alpha_i dx^i dt+\gamma_{\alpha\beta}\sigma^\alpha_i \sigma^\beta_j dx^i dx^j
\end{equation}
with the base invariant 1-forms $\boldsymbol\sigma^\alpha$ defined from
\begin{equation}\label{sigma def}
d \boldsymbol{\sigma}^{\alpha}=C^\alpha_{\beta\gamma}\boldsymbol{\sigma}^\beta\wedge
\boldsymbol{\sigma}^\gamma \Leftrightarrow \sigma^{\alpha}_{i,j} -
\sigma^\alpha_{j,i}=2C^\alpha_{\beta\gamma} \sigma^{\gamma}_{i}
\sigma^\beta_{j}.
\end{equation}

\noindent (Small Latin letters denote world space indices while
small Greek letters count the different basis one-forms; both types
of indices range over the values of 1, 2, 3).

At this stage, we deem it pertinent to give a further explanation of the use of the term Bianchi Types: we mean it to indicate each one of the nine different three
dimensional Lie groups (as in the original Bianchi' s work).
Therefore the (simply transitive) action of the group on a 3-d
hyper-surface can not a priori determine the nature of the
hyper-surface in which it acts. This is determined by the solution
of the Einstein Field Equations (EFE).

The use of \eqref{line element} in the Einstein field equations results in (see, e.g.\cite{Ryan}, \cite{tchris2}):
\begin{equation}\label{quadratic constraint}
E_o\doteq K^{\alpha\beta}K_{\alpha\beta}-K^2- \mathbf{R}=0
\end{equation}
\begin{equation}\label{linear constraint}
E_\alpha\doteq K^\mu_\alpha
C^\epsilon_{\mu\epsilon}-K^\mu_\epsilon C^\epsilon_{\alpha\mu}=0
\end{equation}
\begin{equation}\label{dynamic equation}
\begin{split}
E_{\alpha\beta} & \doteq \dot{K}_{\alpha\beta}+N\left(2K^\tau_\alpha
K_{\tau\beta}-K K_{\alpha\beta}\right)+\\
&+ 2N^\rho\left(K_{\alpha\nu}
C^\nu_{\beta\rho}+K_{\beta\nu} C^\nu_{\alpha\rho} \right)-N
\mathbf{R}_{\alpha\beta}=0
\end{split}
\end{equation}
where
\begin{equation}\label{curvature}
K_{\alpha\beta}=-\frac{1}{2N}\left(\dot{\gamma}_{\alpha\beta}+2\gamma_{\alpha\nu}
C^\nu_{\beta\rho} N^\rho+2\gamma_{\beta\nu} C^\nu_{\alpha\rho}
N^\rho \right)
\end{equation}
is the extrinsic curvature of the three-dimensional hypersurface and
\begin{equation}\label{Ricci}
\begin{array}{cc}
\mathbf{R}_{\alpha\beta}=&C^\kappa_{\sigma\tau} C^\lambda_{\mu\nu}
\gamma_{\kappa\alpha}\gamma_{\beta\lambda}\gamma^{\sigma\nu}\gamma^{\tau\mu}+
2C^\kappa_{\lambda\beta} C^\lambda_{\alpha\kappa}+ 2
C^\mu_{\kappa\alpha}
C^\nu_{\beta\lambda}\gamma_{\mu\nu}\gamma^{\kappa\lambda}+\\
&\\
 &2C^\lambda_{\kappa\beta}
C^\mu_{\mu\nu}\gamma_{\alpha\lambda}\gamma^{\kappa\nu}+ 2
C^\lambda_{\kappa\alpha}
C^\mu_{\mu\nu}\gamma_{\beta\lambda}\gamma^{\kappa\nu}
\end{array}
\end{equation}
is its Ricci tensor.

In \cite{tchris2}, particular spacetime coordinate transformations
have been found, which reveal as symmetries of (\ref{quadratic
constraint}), (\ref{linear constraint}), and (\ref{dynamic
equation}) the following induced transformations of the dependent
variables $N$, $N_{\alpha}$, $\gamma_{\alpha\beta}$:
\begin{equation}\label{gaugetrans}
\tilde{N}=N, \quad
\tilde{N}_{\alpha}=\Lambda^\rho_\alpha\,(N_\rho+\gamma_{\rho\sigma}\,P^\sigma),
\quad
\tilde{\gamma}_{\mu\nu}=\Lambda^\alpha_\mu \, \Lambda^\beta_\nu\gamma_{\alpha\beta}
\end{equation}
where the matrix $\Lambda$ and the triplet $P^\alpha$ must satisfy:
\bsub\label{intcon}\be
 \Lambda^\alpha_\rho \, C^\rho_{\beta\gamma}& =
&C^\alpha_{\mu\nu}\
\Lambda^\mu_\beta \,\Lambda^\nu_\gamma\ \\
2\,P^\mu\,C^\alpha_{\mu\nu}\Lambda^\nu_\beta &=&
\dot{\Lambda}^\alpha_\beta\ee\esub These transformations were first
presented in \cite{Jantzen1}, see also the discussion on p. 3586 of
\cite{tchris2} and $\Lambda, P^\alpha$ describe the action of the
Automorphism group on the various components of the line element.

For all Bianchi Types, this system of equations admits solutions
that contain three arbitrary functions of time plus several
constants depending on the automorphism group of each Type. The
three functions of time are distributed among $\Lambda$ and
$P^\alpha$ (which also contains derivatives of these functions). So
one can use this freedom either to simplify the form of the scale
factor matrix or to set the shift vector $N^\alpha$ to zero. The
second action can always be taken, since, for every Bianchi Type,
all three functions appear in $P^\alpha$.

In this work we adopt the latter point of view. Having used the
freedom stemming from the three arbitrary functions in order to set
the shift vector to zero, there is still a remaining ``gauge''
freedom consisting of a constant $\Lambda^{\alpha}_{\beta}$
(automorphism group matrices of the Lie group defined by the
structure constants $C^{\alpha}_{\beta\gamma}$. Indeed, the system
(\ref{intcon}) accepts the solution $\Lambda^{\alpha}_{\beta} =
\text{const.}$, $P^{\alpha} = \mathbf{0}$. The latter are also known
in the literature as �rigid� symmetries \cite{Coussaert}.

The quadratic constraint \eqref{quadratic constraint} can be used in
order to define the lapse function $N^2$, since it is algebraically
contained in it. In the Type I case, where the quadratic constraint does not determine the lapse, the well known Taub
time gauge $N^2$ equals the determinant of $\gamma_{\alpha\beta}$ can be
used instead. One can further more see, using the definition of
$K_{\alpha\beta}$, that the dynamical equations \eqref{dynamic
equation} involve only $N^2$ and not $N$. Thus $N^2$ is of a nominal
value (not a positive definite function), a fact that will enable
the method to produce \emph{all} hyper-surface homogeneous
spacetimes, even the ones with Euclidean or neutral signature.

The generators of the corresponding motions
$\tilde{\gamma}_{\mu\nu}=\Lambda^\alpha_\mu \, \Lambda^\beta_\nu
\, \gamma_{\alpha\beta}$ induced in the space of the dependent variables spanned by $\gamma_{\alpha\beta}$'s (the lapse is given in terms of $\gamma_{\alpha\beta}$, $\dot{\gamma}_{\alpha\beta}$ by algebraically solving the quadratic constraint equation), are:
\begin{equation}\label{genX}
X_{I}=\lambda^\rho_{I\alpha}\,\gamma_{\rho\beta}\,
\frac{\partial}{\partial\gamma_{\alpha\beta}}
\end{equation}
with $\lambda^\alpha_\beta$ satisfying
\begin{equation}\label{deflamda}
\lambda^\alpha_{I\rho}\,C^\rho_{\beta\gamma}=\lambda^\rho_{I\beta}\,C^\alpha_{\rho\gamma}+
\lambda^\rho_{I\gamma}\,C^\alpha_{\beta\rho}.
\end{equation}

These generators define a Lie algebra and each one of them induces, through its integral curves, a transformation on the configuration space spanned by the $\gamma_{\alpha\beta}$'s \cite{tchris4}. If a generator is brought to its normal form (i.e., $\frac{\partial}{\partial z_i}$), then the Einstein field equations, written in terms of the new dependent variables, will not explicitly involve $z_i$. They thus become a first order system in the function $\dot{z}_i$ \cite{Stephani}. If the aforesaid Lie algebra is abelian, then all generators can be brought to their normal form simultaneously. If the Lie algebra is non-abelian, then we can diagonalize in one step those generators corresponding to any eventual abelian subgroup. The rest of the generators (not brought in their normal form) continue to define a symmetry of the reduced system of the field equations if the Lie algebra of the $X_{(I)}$'s is solvable \cite{Olver}. One can thus repeat the previous step by choosing one of these remaining generators and bring it to its normal form. This choice will of course depend upon the simplifications brought to the system at the previous level. Finally, if the Lie algebra does not contain any abelian subgroup, one can always choose one of the generators, bring it to its normal form, reduce the system of equations, and search for its symmetries (if there are any).
Lastly, two further symmetries of (\ref{quadratic constraint}), (\ref{linear constraint}), and (\ref{dynamic equation}) are also present and can be used in conjunction with the constant automorphisms: The time reparametrization $t \rightarrow t+\alpha$, owing to the non-appearance of time in these equations (the system being autonomous), and the scaling by a constant $\gamma_{\alpha\beta}
\rightarrow \lambda \gamma_{\alpha\beta}$ (homothety) as can be straightforwardly verified. Hence, in every Bianchi Type there are, added to the $X_{(I)}$ generators, also the following generators:
\begin{equation}\label{Y1}
 Y_1  = \frac{\partial}{\partial t}
\end{equation}
\begin{equation}\label{Y2}
Y_2=\gamma_{11}\frac{\partial}{\partial\gamma_{11}}+
\gamma_{12}\frac{\partial}{\partial\gamma_{12}}+
\gamma_{13}\frac{\partial}{\partial\gamma_{13}}+
\gamma_{22}\frac{\partial}{\partial\gamma_{22}}+
\gamma_{23}\frac{\partial}{\partial\gamma_{23}}+
\gamma_{33}\frac{\partial}{\partial\gamma_{33}}
 \end{equation}
These generators commute among themselves, as well as with the $X_{(I)}$'s:
\begin{equation}
\left[X_I,Y_\alpha\right]=0 \quad\quad \left\{ I=1,2,3,4
\, | \, \alpha=1,2\right\}
\end{equation}

\section{Bianchi Type $IV$}
For this type the structure constants are
\begin{equation}\label{�������� �����}
\begin{array}{lll}
C^1_{\ph{1}13}=-C^1_{\ph{1}31}=C^2_{\ph{2}23}=-C^2_{\ph{2}32}=\frac{1}{2}\\
C^1_{\ph{1}23}=-C^1_{\ph{1}32}=\frac{1}{2}\\
C^\alpha_{\ph{a}\beta\gamma}=0 & for\, all\, other\, values\, of\,
\alpha \beta \gamma
\end{array}
\end{equation}
Using these values in the defining relation (\ref{sigma def}) of the
1-forms $\sigma^{\alpha}_{i}$ we obtain
\begin{equation}
\sigma^\alpha_i=\left(\begin{matrix} 0 & e^{-x} &
x\,e^{-x} \cr 0 & 0 & -e^{-x} \cr
1 & 0 & 0
\end{matrix}
\right)
\end{equation}
The corresponding vector fields $\xi^i_\alpha$ (satisfying
$\left[\xi_\alpha,\xi_\beta\right]=C^\gamma_{\ph{a}\alpha\beta}
\xi_\gamma$) with respect to which the Lie Derivative of the above
1-forms is zero are: \be\label{killing}
\begin{array}{lll}
\xi_1=-\partial_y & \xi_2=\partial_z &
\xi_3=\partial_x+(y-z)\partial_y+z\partial_z
\end{array}
\ee

The Time Depended A.I.D.'s are described by
\begin{equation}\label{Aut �}
 \Lambda^\alpha_{\ph{a}\beta}=
\left(\begin{matrix} P(t) & P(t)\ln(c\,
P(t)) & x(t) \cr
  0 & P(t) & y(t)
  \cr
  0 & 0 & 1\end{matrix}
 \right)
\end{equation} and
\be \label{Aut P} P^\alpha
&=&\left(x(t)\,\left(\ln\frac{x(t)}{P(t)}\right)'-y'(t),y(t)\,\left(\ln\frac{y(t)}{P(t)}\right)',-\left(\ln P(t)\right)'\right) \ee where $P(t), x(t)$ and $y(t)$ are
arbitrary functions of time. As we have already remarked the three
arbitrary functions appear in $P^\alpha$ and thus can be used to set
the shift vector to zero.

The remaining symmetry of the EFE's is, consequently, described by
the constant matrix:
\begin{equation}\label{Outer Aut}
M=\left(\begin{matrix}e^{s_{1}}& s_2 &s_{3} \cr 0
&e^{s_{1}}&s_{4} \cr 0&0&1\end{matrix}\right)
\end{equation}
where the parametrization has been chosen so that the matrix becomes
identity for the zero value of all parameters.

 Thus the induced transformation on the scale factor matrix is
$\tilde{\gamma}_{\alpha\beta}=M^{\mu}_{\alpha}M^{\nu}_{\beta}\gamma_{\mu\nu}$,
which explicitly reads:
\begin{equation}\label{gamma new}
\left\{
\begin{array}{l}
\tilde{\gamma}_{11}=e^{2\,{s_1}}\,{{\gamma }_{11}}\\\
\\
\tilde{\gamma}_{12}=e^{{s_1}}\,s_2\,{{\gamma }_{12}}+e^{2\,s_1}\,\gamma_{12}\\
\\
\tilde{\gamma}_{13}=e^{{s_1}}\,\left( {s_3}\,{{\gamma }_{11}} + {s_4}\,{{\gamma }_{12}} + {{\gamma }_{13}}\right)\\
\\
\tilde{\gamma}_{22}=e^{2\,s_1}\,\gamma_{22}+2\,e^{s_1}\,s_2\,\gamma_{12}+s_2^2\,\gamma_{11}\\
\\
\tilde{\gamma}_{23}=e^{{s_1}}\,\left( {s_3}\,{{\gamma }_{12}} +s_4\,\gamma_{22}+ {{\gamma }_{23}}\right)+
s_2\,\left(s_3\,\gamma_{11}+\gamma_{13}+s_4\,\gamma_{12}\right)\\
\\
\tilde{\gamma}_{33}={{s_3}}^2\,{{\gamma }_{11}} + {s_3}\,
   \left({{\gamma }_{12}} + {{\gamma }_{13}} \right)  + {{s_4}}^2\,{{\gamma }_{22}} +
  2\,{s_4}\,\left(s_3\,\gamma_{12}+{{\gamma }_{23}}\right) + {{\gamma }_{33}}
\end{array}\right.
\end{equation}

The previous equations, define a group of transformations $G_{r}$ of
dimension  $r=dim(Aut(IV))=4$. The four generators of the group are:
\begin{equation}\label{X1}
X_{1}=2\,\gamma_{11}\frac{\partial}{\partial\gamma_{11}}+2\,\gamma_{12}\frac{\partial}{\partial\gamma_{12}}
+\gamma_{13}\frac{\partial}{\partial\gamma_{13}}+2\,\gamma_{22}\frac{\partial}{\gamma_{22}}+
\gamma_{23}\frac{\partial}{\gamma_{23}}
\end{equation}
\begin{equation}\label{X2}
X_{2}=\gamma_{11}\frac{\partial}{\partial\gamma_{12}}
+2\,\frac{\partial}{\partial\gamma_{22}}
+\gamma_{13}\frac{\partial}{\partial\gamma_{23}}
\end{equation}
\begin{equation}\label{X3}
X_{3}=\gamma_{11}\frac{\partial}{\partial\gamma_{13}}
+\gamma_{12}\frac{\partial}{\partial\gamma_{23}}
+2\,\gamma_{13}\frac{\partial}{\partial\gamma_{33}}
\end{equation}
\begin{equation}\label{X4}
X_{4}=\gamma_{12}\frac{\partial}{\partial\gamma_{13}}
+\gamma_{22}\frac{\partial}{\partial\gamma_{23}}
+2\,\gamma_{23}\frac{\partial}{\partial\gamma_{33}}
\end{equation}

The algebra $\textsl{g}_{r}$ that corresponds to the group $G_{r}$
has the following table of  non-zero commutators:
\begin{equation}\label{���������}
\begin{array}{lll}
\left[X_{1},X_{2}\right]=0,&\left[X_{1},X_{3}\right]=X_3,&\left[X_{1},X_{4}\right]=X_{4},
\\ \left[X_{2},X_{3}\right]=0,& \left[X_{2},X_{4}\right]=X_3,&\left[X_{3},X_{4}\right]=0
\end{array}
\end{equation}

As it is evident from the above commutators (\ref{���������})  the
group is non-abelian, so we cannot diagonalize  at the same time all
the generators. However, if we calculate the derived algebra of
$\textsl{g}_{r}$, we have
\begin{equation}
\textsl{g}_{r'}=\left\{[X_{A},X_{B}]: X_{A}, X_{B}\in
\textsl{g}_{r}\right\}\Rightarrow
\textsl{g}_{r'}=\left\{X_{3},X_{4}\right\}
\end{equation}
and furthermore, it's second derived algebra reads:
\begin{equation}
\textsl{g}_{r''}=\left\{[X_{A},X_{B}]: X_{A}, X_{B}\in
\textsl{g}_{r'}\right\}\Rightarrow \textsl{g}_{r''}=\left\{0\right\}
\end{equation}

Thus, the group  $G_{r}$ is solvable since the $\textsl{g}_{r''}$ is
zero. As it is evident  $X_{3}, X_{4}, Y_{2}$ generate an Abelian
subgroup, and we can, therefore,  bring them to their normal form
simultaneously. The appropriate transformation of the dependent
variables is: \be\label{gammav} \left\{
\begin{array}{l}
\gamma_{11}= e^{{u_1}} \\
\\
\gamma_{12}=e^{{u_1}}\,u_2\\
\\
\gamma_{13}=e^{{u_1}}\,\left(-e^{u_4}\,u_2^2 + u_3+u_2\,u_5 \right) \\
\\
\gamma_{22}=e^{{u_1}+{u_4}}\\
\\
\gamma_{23}= e^{{u_1}}\,\left( u_2\,(u_3-1) +e^{u_4}\,u_5 \right) \\
\\
\gamma_{33}=e^{u_1}\,\left(e^{-2\,u_6}+{{u_3}}^2 -e^{-u_4}\,u_2^2\,(2\,u_3-1)+2\,u_2\,\,u_5\,(u_3-1)
+e^{u_4}\,u_5^2 \right)
\end{array}\right.
\ee

In these coordinates the generators $Y_{2}, X_{A}$  assume the form:

\be\label{generatorsv}
\begin{array}{lll}
Y_2=\frac{\partial}{\partial u_1} \quad X_4=\frac{\partial}{\partial
u_5} \quad X_3=\frac{\partial}{\partial u_3}  \\
\\
X_2=\frac{\partial}{\partial u_2}+
\left(e^{-u_4}\,u_2-u_5\right)\,\frac{\partial}{\partial
u_3}+2\,e^{-u_4}\,u_2\,\frac{\partial}{\partial
u_4}+e^{-2\,u_4}\,\left(e^{u_4}-2\,u_2^2\right)\,\frac{\partial}{\partial
u_5}\\
\\
X_1 =2\,\frac{\partial}{\partial u_1}-u_3\,\frac{\partial}{\partial u_3}+\left(e^{-u_4}\,u_2-u_5\right)\,
\frac{\partial}{\partial u_5}+\frac{\partial}{\partial u_6}
\end{array}
\ee

Evidently, a first look at (\ref{gammav}) gives the feeling that it
would be hopeless even to write down the Einstein equation. However,
the simple form of the first three of the generators
(\ref{generatorsv}) ensures us that these equations will be of first
order in the functions $\dot{u}_1$,  $\dot{u}_3$ and $\dot{u}_5$.
\subsection{Description of the Solution Space}

Before we begin solving the Einstein equations, a few comments for
the possible values of the functions $u_{i}, i={1,\ldots,6}$ will
prove very useful.

The determinant of  $\gamma_{\alpha\beta}$, is
\begin{equation}\label{detgamma}
det[\gamma_{\alpha\beta}]=e^{3\,u_1 - 2\,u_6}\,\left(e^{u_4}-u_2^2\right)
\end{equation}
so we must have $e^{u_4}\,>\,u_2^2$ .

The two linear constraint equations, written in the new variables
(\ref{gammav}), give \be E_1=0 \Rightarrow
e^{u_4}\,\dot{u}_3+u_2^2\,\dot{u}_4+u_2\,(e^{u_4}\,\dot{u}_5-\dot{u}_2)=0 \ee
\be E_2=0\Rightarrow(3\,e^{u_4}+u_2)\,\dot{u}_2-
e^{u_4}\,(1+3\,u_2)\,\dot{u}_3-(u_2+3\,e^{u_4})\,(u_2\,\dot{u}_4+e^{u_4}\,\dot{u}_5)=0 \ee
Solving this system for the functions $\dot{u}_3,\dot{u}_5$ we have
\be \dot{u}_3=0,& \dot{u}_5=e^{-u_4}\,\left(\dot{u}_2-u_2\,\dot{u}_4\right)\ee yielding to
\be u_3=k_3, & u_5=k_5+e^{-u_4}\,u_2\ee
Now, these values of $u_3,u_5$
make $\gamma_{13},\gamma_{23}$ functionally dependent upon
$\gamma_{11},\gamma_{12},\gamma_{22}$ (see (\ref{gammav})). It is
thus possible to set these two components to zero by means of an
appropriate constant automorphism.

\emph{Without loss of generality, we can start our investigation of
the solution space for Type $IV$ vacuum Bianchi Cosmology from a
block-diagonal form of the scale-factor matrix (and, of course, zero
shift)} \be\label{gammau1} \gamma_{\alpha\beta}=\left(
\begin{matrix}
\gamma_{11} & \gamma_{12} & 0 \cr \gamma_{12} & \gamma_{22} & 0 \cr
0 & 0 & \gamma_{33}
\end{matrix}
\right) \ee Note that this conclusion could have not been reached
off mass-shell, due to the fact that the time-dependent Automorphism
(\ref{Aut �}) does not contain the necessary two arbitrary functions
of time in the (13) and (23) components (besides the fact that all
the freedom in arbitrary functions of time has been used to set the
shift to zero). As we have earlier remarked, since the algebra
(\ref{���������}) is solvable, the remaining (reduced) generators
$X_1,X_2$ (corresponding to diagonal constant automorphisms) as well
as $Y_2$ continue to define a Lie-Point symmetry of the reduced
EFE's and can thus be used for further integration of this system of
equations.

The remaining (reduced) automorphism generators are \be
X_{1}=\gamma_{11}\,\frac{\partial}{\partial\gamma_{11}}
+\gamma_{12}\,\frac{\partial}{\partial\gamma_{12}}
+\gamma_{22}\,\frac{\partial}{\partial\gamma_{22}}\non\\
X_{2}=\gamma_{11}\,\frac{\partial}{\partial\gamma_{12}}
+2\,\gamma_{12}\,\frac{\partial}{\partial\gamma_{22}} \ee
The appropriate change of dependent variables which brings these
generators -along with $Y_2$- into normal form, is described by the
following scale-factor matrix :\be \gamma_{\alpha\beta}=\left(
\begin{matrix}
e^{u_1+2\,u_6} &
e^{u_1+2\,u_6}\,u_2 & 0 \cr e^{u_1+2\,u_6}\,u_2 & e^{u_1+2\,u_6}\,\left(u_2^2+u_4\right) & 0 \cr 0 & 0 & e^{u_1}
\end{matrix} \right)
\ee The generators are now reduced to \be
Y_2=\frac{\partial}{\partial u_1}, \, X_2=\frac{\partial}{\partial
u_2}, \, X_1=\frac{\partial}{\partial u_6} \ee indicating that the
system will be of first order in the derivatives of these variables.
The remaining variable $u_4$ will enter, (along with
$\dot{u}_4,\,\ddot{u}_4$) explicitly in the system and is therefore
advisable (if not mandatory) to be used as the time parameter, i.e.
to effect the change of time coordinate \be t\rightarrow u_4(t)=s,
\, u_1(t)\rightarrow u_1(t(s)), \, u_2(t)\rightarrow u_2(t(s)), \,
u_6(t)\rightarrow u_6(t(s)). \ee This choice of time will of course
be valid only if $u_4$ is not a constant. We are thus led to
consider two cases according to the constancy or non-constancy of
this variable.

\subsubsection{Case I: $u_4(t)\,=\,k_4$}
In these variables the first two linear constraint equations are
identically satisfied, and the determinant of
$\gamma_{\alpha\beta}$, is
\begin{equation*}
det[\gamma_{\alpha\beta}]=e^{3\,u_1 + 4\,u_6}\,k_4
\end{equation*}
so we must have $k_4>0$ . The third linear
constraint reads \be E_3=0\Rightarrow
\dot{u}_2+4\,k_4\,\dot{u}_6=0\Rightarrow u_2=k_2-4\,k_4\,u_6\ee  Substituting
this value into the quadratic constraint equation $E_0$ we obtain
for the lapse function
\be N^2=\frac{k_4}{12\,k_4+1}\,e^{u_1}\,(3\,\dot{u}_1^2+8\,\dot{u}_1\,\dot{u}_6-4\,(4\,k_4-1)\,\dot{u}_6^2)\ee
If we substitute this value of the lapse function into the equations of motion \eqref{dynamic equation}, we are left with the unknown functions $u_1,u_6$. The strategy we follow is to solve one of \eqref{dynamic equation} for a second derivative of some function, say $\ddot{u}_6$ and replace the result into the rest of the equations. In order to do that, we must ensure that the coefficient of $\ddot{u}_6$ does not equals to zero. Looking at $E_{11}=0$ the coefficient of $\ddot{u}_6$ is proportional to
\be\label{coeff u6} \dot{u}_1\,(\dot{u}_1+2\,(4\,k_4+1)\,\dot{u}_6)\Rightarrow \left\{u_1=k_1\quad\text{or}\quad u_1=k_1-2\,u_6-8\,k_4\,u_6\right\}\non\ee
Thus we are forced to examine the above equalities before we solve $E_{11}=0$, for $\ddot{u}_6$.

The first possibility $u_1=k_1$ yields an inconsistency, so we are left with the second, i.e.
\be u_1=k_1-2\,u_6-8\,k_4\,u_6\ee
The above choice satisfies \emph{all} the spatial equations $E_{\alpha\beta}=0$ and gives the lapse function
\be N^2=16\,e^{k_1-2(4\,k_4+1)\,u_6}\,k_4^2\,\dot{u}_6^2\ee

Redefining the constants $k_1=\ln\kappa^2,\,k_4=\frac{-\mu}{4\,(\mu-1)}$, choosing a time  parametrization
$u_6=(\mu-1)\,\ln(t)$, and using the
automorphism matrix  \eqref{Outer Aut} with entries
$s_1=0,\,s_2=-k_2,\,s_3=0,\,s_4=0$ we arrive
at the line element
\begin{multline}\label{tsoubelis}d\,s^2=-\mu^2\,(\bs{d}\,t)^2+t^{2\,\mu}\,(\bs{\sigma^1})^2+2\,t^{2\,\mu}\,\mu\,\ln t\,\bs{\sigma^1}\,\bs{\sigma^2} \\+t^{2\,\mu}\,\mu\,\left(\mu\,\ln^2t-\frac{1}{4\,(\mu-1)}\right)\,(\bs{\sigma^2})^2+t^2\,(\bs{\sigma^3})^2 \end{multline}
with $t>0,\,0<\mu<1$. In the above line element we have dropped the
constant $\kappa$ since this line element admits a homothetic vector
field \be\label{hom diagonal}
H=\partial_t+\left(\mu\,z-y\,(\mu-1)\right)\,\partial_y-z(\mu-1)\,\partial_z\ee
Line element \eqref{tsoubelis} was first derived by Harvey and
Tsoubelis \cite{Tsoubelis} and admits, besides the three Killing
fields \eqref{killing} three more, namely
\begin{align}
\xi_4=&e^{-x/\mu}\,\partial_t+\frac{e^{-x/\mu}}{t}\,\partial_x \\
\xi_5=&y\,e^{-x/\mu}\,\partial_t+\frac{y\,e^{-x/\mu}}{t}\,\partial_x+f_1\,\partial_y+f_2\,\partial_z\\
\xi_6=&z\,e^{-x/\mu}\,\partial_t+\frac{z\,e^{-x/\mu}}{t}\,\partial_x+f_2\,\partial_y+f_3\,\partial_z
\end{align}

with
\begin{align}f_1=&\frac{\mu\,e^{(2\,\mu-1)x/\mu}\,t^{-2\,\mu+1}}{(-2\,\mu+1)^3}\bigg(4\,\mu^2\,(2\,\mu-1)^2\,(\mu-1)\,\ln^2t \non\\
&-8\,\mu\,(\mu-1)\,(2\,\mu-1)(-\mu+(2\mu-1)\,x)\,\ln t\non\\
&-\mu+4\,(\mu-1)\,(\mu+(1-2\,\mu)\,x)^2\bigg)\non\\
f_2=&\frac{4\,\mu\,(\mu-1)\,e^{(2\,\mu-1)x/\mu}\,t^{-2\,\mu+1}}{(-2\,\mu+1)^2}\big(\mu\,(2\,\mu-1)\,\ln t+\mu+(1-2\,\mu)\,x\big)\non\\ f_3=&\frac{4\,\mu\,(\mu-1)\,e^{(2\,\mu-1)x/\mu}\,t^{-2\,\mu+1}}{-2\,\mu+1}\end{align}
There is thus a $G_6$ symmetry group acting (of course, multiply
transitively) on each $V_3$ of this metric. However, it is
interesting to note that we have not imposed the extra symmetry from
the beginning, but rather it emerged as a result of the
investigation process.

Having ensured that the coefficient of $\ddot{u}_6$ at $E_{11}=0$ is not zero, we can solve this equation for $\ddot{u}_6$ and insert the result into the rest of the spatial equations. But doing that we end up with a zero lapse, indicating that the only solution for this case is described by the line element \eqref{tsoubelis}.

\subsubsection{Case II: $u_4(t)\,=\,t$}
In this case the determinant of the scale factor matrix equals to
\be det[\gamma_{\alpha\beta}]=e^{3\,u_1+4\,u_6}\,t \non\ee so we
must demand that $t>0$ in order for $\gamma_{\alpha\beta}$ to be
positive defined.

The first two linear constraints are identically zero while the
third one $E_3=0$ can be used to define the function $u_2$ \be
E_3=0\Rightarrow \dot{u}_2+4\,t\,\dot{u}_6+1=0 \Rightarrow
u_2=k_2-t-4\,\int t\,\dot{u}_6\,d\,t \ee and the quadratic
constraint $E_o=0$ defines the lapse function $N^2$ \be\label{lapse
IV}N^2=\frac{e^{u_1}}{12\,t+1}
\,\Big(3\,t\,\dot{u}_1^2+8\,t\,\dot{u_1}\,\dot{u}_6
-4\,t\,(4\,t-1)\,\dot{u}_6^2-2\,(4\,t-1)\,\dot{u}_6+2\,\dot{u}_1-1\Big)
\ee Substituting the above values of the lapse $N^2$ and the
function $u_2$ in equation $E_{33}=0$ we find the coefficient of
$\ddot{u}_6$ is proportional to \be
\dot{u}_1\,\left(4\,t\,\dot{u}_1-(4\,t-1)\,(4\,t\,\dot{u}_6+1)\right)\non\ee
a quantity that can be safely regarded different from zero, since
it's nihilism leads either to zero lapse or to inconsistency when is
combined with the rest of the dynamical equations. Thus we can solve
$E_{33}=0$ for $\ddot{u}_6$ and substitute it to $E_{11}=0$. In
order to solve this equation for $\ddot{u}_1$ we must be assured
that it's coefficient does not equals to zero. Setting this
coefficient equal to zero we arrive to the following equation \be
\dot{u}_1=1+(4\,t-1)\,\dot{u}_6\non\ee which is unacceptable because
it leads to inconsistency. After solving equation $E_{11}=0$ for
$\ddot{u}_1$ we finally arrive to the following polynomial system of
first order in $\dot{u}_1,\,\dot{u}_6$ \be\label{d2du1,
d2du6}\ddot{u}_1=\bra{\dot{u}_1}\,B_1\,\ket{\dot{u}_6}, &
\ddot{u}_6=\bra{\dot{u}_1}\,B_2\,\ket{\dot{u}_6}\ee where we have
used  the notation $\bra{\dot{u}_i}=\left( 1 \, \dot{u}_i \,
\dot{u}_i^2\, \dot{u}_i^3 \right)$ and
$\ket{\dot{u}_i}=\bra{\dot{u}_i}^t$ with the $4\times4$ matrices
$B_1,\,B_2$ given by
\be\label{B1} B_1&=\frac{1}{t\,(12\,t+1)}\begin{pmatrix} -4\,t-1 & -32\,t^3+2\,t & -64\,t^3+4\,t & 0 \\
24\,t^2+1 & 16\,t^2+4\,t & 8\,t^2\,(4\,t+1)\,(12\,t-1) & 0 \\
12\,t^2-t & -16\,t^2 & 0 & 0 \\
-6\,t^2& 0 & 0 & 0\end{pmatrix}\non\ee
\be\label{B2}B_2&=\frac{1}{2\,t\,(12\,t+1)}\begin{pmatrix} 12\,t+3 & 144\,t^2+8\,t-6 & 16\,t\,(36\,t^2+7\,t-1) & 16\,t^2\,(48\,t^2+8\,t-1)\\
-4 & -24\,t & -32\,t^2 & 0 \\
-6\,t & -12\,t^2 & 0 & 0 \\
0& 0 & 0 & 0
\end{pmatrix} \non \ee
Due to the form of $B_1, B_2$
(their components are rational functions of the time $t$), system
\eqref{d2du1, d2du6} can be partially integrated with the help of the
following Lie-B\"{a}cklund transformation
\bsub\label{du1, du6} \be \dot{u}_1&=&\frac{(48\,t^2+16\,t+1)\,\dot{r}(t)-2\,(12\,t-1)\,\tan r(t)}{4\,\sqrt{t}\,(12\,t+1)}\\ \non\\
\dot{u}_6&=& -\frac{\sqrt{t}\,(12\,t+1)\,\dot{r}(t)+6\,\sqrt{t}\,\tan r(t)+24\,t+2}{8\,t\,(12\,t+1)}\ee\esub
yielding the single second order ODE for the function $r(t)$
\begin{multline}\label{equation r II}
\ddot{r}=\left(\tan r+\sqrt{t}\right)\,\dot{r}^2-2\,\frac{(6\,t+1)\,\tan r+6\,\sqrt{t}}{\sqrt{t}\,(12\,t+1)}\,\dot{r}\\+
\frac{36\,t^2\,\tan^2 r+36\,\sqrt{t^3}\,\tan r-12\,t-1}{\sqrt{t^3}\,(12\,t+1)^2}\end{multline}
This equation contains all the information concerning the
unknown part of the solution space of the Type $IV$ vacuum
Cosmology. Unfortunately, it does not posses any Lie-point
symmetries that can be used to reduce its order and ultimately
solve it. However, its form can be substantially simplified
through the use of new dependent and independent variable
$(\rho,u(\rho))$ according to  $
r(t)=\mp\arcsin{\frac{u(\rho)}{\sqrt{\rho^2-1}}},\,
t=\frac{1}{6\,(\rho-1)} $ thereby
obtaining the equation \be\label{final u IV}
  \ddot{u}=\pm\frac{1-\dot{u}^2}{\sqrt{6\,(\rho-1)\,(\rho^2-u^2-1)}}
  \Rightarrow \ddot{u}^2=\frac{(1-\dot{u}^2)^2}{6\,
  (\rho-1)\,(\rho^2-u^2-1)}
  \ee

This equation is a special case of the general equation
\be\label{equ general} \ddot{u}^2=\frac{(1-\dot{u}^2)^2}{(\kappa+ \lambda\,\rho)\,(\rho^2-u^2-1)}\ee with the values $\kappa=-6,\, \lambda=6$. The general solution of \eqref{equ general} was first given in \cite{ChrTer CQG} and can be obtained as follows: We first apply the
contact transformation: \be\label{contact II}
\begin{split}
u(\rho)& =
-\frac{8}{\lambda}\,y(\xi)+\frac{4\,(2\,\xi-1)}{\lambda}\,y'(\xi) &\rho & =  -\frac{\kappa}{\lambda}+\frac{4}{\lambda}\,y'(\xi)\\
\dot{u}(\rho)& =  2\,\xi-1 & \ddot{u}(\rho)& = \frac{\lambda}{2\,y''(\xi)}
\end{split}\ee
which reduces it to \be\label{y equation} \xi^2\,(\xi-1)^2\,{y''}^2=
-4y'\,(\xi\,y'-y)^2+4\,{y'}^2\,(\xi\,y'-y)-\frac{\kappa}{2}\,{y'}^2+
\frac{\kappa^2-\lambda^2}{16}\,y'
 \ee

This equation is a special form of the equation SD-Ia, appearing in \cite{Cosgrove}, where  a classification of all second order second
degree ordinary differential equations was performed. The general
solution of (\ref{y equation}) is obtained with the help of the
sixth Painlev\'{e} transcendent
$w:=\mathbf{P_{VI}}(\alpha,\beta,\gamma,\delta)$ and reads:
\be\label{solution y} y & = &
\frac{\xi^2\,(\xi-1)^2}{4\,w\,(w-1)(w-\xi)}\,\left(w'-\frac{w\,(w-1)}{\xi\,(\xi-1)}\right)^2
\nonumber \\ & & +\frac{1}{8}\,(1\pm
\sqrt{2\,\alpha})^2\,(1-2\,w)-\frac{\beta}{4}\,\left(1-\frac{2\,\xi}{w}\right)
\nonumber \\ & &
-\frac{\gamma}{4}\,\left(1-\frac{2\,(\xi-1)}{w-1}\right)+ \left(\frac{1}{8}-\frac{\delta}{4}\right)
\, \left(1-\frac{2\,\xi\,(w-1)}{w-\xi}\right) \ee where the sixth
Painlev\'{e} transcendent
$w:=\mathbf{P_{VI}}(\alpha,\beta,\gamma,\delta)$ is defined by the
ODE:

\be\label{Painleve 6} w'' & = & \frac{1}{2}\left( \frac{1}{w-1} +
\frac{1}{w} + \frac{1}{w-\xi } \right) \,{w'}^2 -\left( \frac{1}{\xi-1} + \frac{1}{\xi} + \frac{1}{w-\xi}
\right) \,w' \nonumber \\
& & +\frac{w\,\left( w-1 \right) \,\left( w-\xi\right) }
  {{\xi^2\,\left( \xi-1 \right) }^2} \left(\alpha +\beta\,
  \frac{\xi}{{w}^2}  + \gamma\, \frac{\left(\xi-1 \right)}
  {{\left( w-1 \right) }^2} +\delta\,\frac{\xi\,\left(\xi -1\right) }
   {{\left( w-\xi\right) }^2}\right) \ee

 The values of the parameters
$\left(\alpha,\beta,\gamma,\delta\right)$ of the Painlev\'{e}
transcendent, can be obtained from the solution of the following
system: \bsub\label{system} \be \alpha-\beta+\gamma-\delta \pm
\sqrt{2\,\alpha}+1 & =&-\frac{\kappa}{2} \\
\left(\beta+\gamma\right)\,\left(\alpha+\delta \pm \sqrt{2\,\alpha}\right) &=&0 \\
\left(\gamma-\beta\right)\,\left(\alpha-\delta \pm
\sqrt{2\,\alpha}+1\right)+\frac{1}{4}\,\left(\alpha-\beta-\gamma+\delta
\pm \sqrt{2\,\alpha}\right)^2 & = & \frac{\kappa^2-\lambda^2}{16} \\
\frac{1}{4}\,\left(\gamma-\beta\right)\,\left(\alpha+\delta \pm
\sqrt{2\,\alpha}\right)^2+\frac{1}{4}\,\left(\beta+\gamma\right)^2\,\left(\alpha-\delta
\pm \sqrt{2\,\alpha}+1\right) & = & 0 \ee \esub

Inserting in (\ref{system}) the values of $\kappa=-6, \lambda=6$ for
Type $IV$, we have twenty-four solutions (counting multiplicities) of this system. In order for the parameters $(\alpha,\beta,\gamma,\delta)$ to be real numbers we end up only with three possibilities \bsub\label{par}\be (\alpha,\beta,\gamma,\delta)&=& \left(\disfrac{1}{2},-\disfrac{3}{2},\disfrac{3}{2}, \disfrac{1}{2}\right)\\
(\alpha,\beta,\gamma,\delta)&=& \left(2+\sqrt{3},0,0, -1\right)\\
(\alpha,\beta,\gamma,\delta)&=& \left(2-\sqrt{3}, 0,0,-1\right)\ee\esub

Gathering all the pieces the final form of the \emph{general} line element describing  Bianchi Type $IV$ vacuum Cosmology is
\be\label{final IV}
d\,s^2&=&\kappa^2\Big(-\frac{e^{u_1(\xi)}}{4\,\xi\,(\xi-1)}\,(\bs{d}\,\xi)^2+ \sqrt{\xi\,(\xi-1)\,y'(\xi)}\,(\bs{\sigma}^1)^2 +2\,\sqrt{\xi\,(\xi-1)\,y'(\xi)}\,u_2(\xi)\,\bs{\sigma}^1\,\bs{\sigma}^2 \non \\&&+ \frac{1}{4}\sqrt{\frac{\xi\,(\xi-1)}{y'(\xi)}}\left(4\,u_2^2(\xi)\,y'(\xi)+1\right)\,(\bs{\sigma}^2)^2+ e^{u_1(\xi)}\,(\bs{\sigma}^3)^2\Big)
\ee
where \be\label{u1 and u2} u_1'(\xi) =  \frac{ 1 - 2\,\xi - 2\,y(\xi )}{2\,\xi\,\left(\xi-1\right)},&
u_2'(\xi) = \disfrac{y(\xi )}{2\,\xi\,y'(\xi)\,\left(\xi-1\right) }\ee
and $ y(\xi)$ is given by \eqref{solution y}.

The above line element does not admit a homothetic field, thus the constant $\kappa$ is essential and together with the two constants of integration inherent in $\bs{P_{VI}}$ indicate the generality of the solution see e.g. \cite{Stephani1}.

\section{Solution space for Types $I,II,V$}
In this section we summarize the results of applying the method in Bianchi Types $I,II,V$. These Types can be considered as "easier" than the rest, a characterization which can be supported both by the time of the discovery of their solution (Kasner (1921), Taub (1951), Joseph (1966)) and the fact that their generalizations with a cosmological constant is also known. Despite all these, the reproduction of all the above metrics by a single method is, in our opinion, worth reporting even in a "telegraphic" sort of presentation. Moreover, along with the known solutions, three new metrics are given; one of Euclidean, one of neutral and one of Lorentzian signature are obtained. The detailed calculations can be found in \cite{unp}.

\subsection{Bianchi Type $I$}
In this model the structure constants, the basis 1-forms and the Killing fields are
\bsub
\begin{equation}
C^\alpha_{\ph{a}\beta\gamma}=0 \qquad \text{for every value of}\quad \alpha, \,\beta, \,\gamma
\end{equation}
\begin{equation}
\boldsymbol{\sigma}^1=\mathbf{d}\,x,\quad
\boldsymbol{\sigma}^2=\mathbf{d}\,y, \quad
\boldsymbol{\sigma}^3=\mathbf{d}\,z
\end{equation}
\begin{equation}\label{killing type I}
\xi_1=\partial_x,\quad
\xi_2=\partial_y,\quad
\xi_3=\partial_z
\end{equation}
\esub
The ensuing metrics are:
\begin{itemize}
\item \textbf{Kasner metrics}. \be\label{Kasner metric}
d\,s^2=-e^{(1+\alpha+\beta)\,\tau}\,d\,\tau^2+e^{\tau}\,d\,x^2+e^{\alpha\,\tau}\,d\,y^2+e^{\beta\,\tau}\,d\,z^2
\ee
where the constants $(\alpha,\beta)$ satisfy $\alpha+\beta+\alpha\,\beta=0$.
The above metric was first given, although in a different form,
in \cite{Kasner} and admits a homothety \be
H=2\,(\alpha+1)\,\partial_\tau+\alpha^2\,x\,\partial_x+y\,\partial_y+(\alpha+1)^2\,z\,\partial_z.
\ee
The metric is particularly interesting for
the values $(\alpha,\beta)=(1,-1/2)\, \text{or}\,
(\alpha,\beta)=(-1/2,1)$: in addition to the three Killing fields
\eqref{killing type I}, there is a fourth of the form \be
\xi_4=y\,\partial_x-x\,\partial_y. \ee The pair of
values $(\alpha,\beta)=(0,0)$ leads the metric \eqref{Kasner metric}
to the standard Minkowski form.
\item \textbf{Harrison metrics}.
\be\label{Kasner II}
d\,s^2 &= e^{(2\,\lambda+\beta^{-1})\,\tau}\,d\,\tau^2+e^{\tau/\beta}\,d\,x^2-e^{\lambda\,\tau}\,\sin\tau\,d\,y^2+\non\\
&+2\,e^{\lambda\,\tau}\,\cos\tau\,d\,y\,d\,z+e^{\lambda\,\tau}\,\sin\tau\,d\,z^2
\ee
which possesses a homothety produced by the field \be
H=-4\lambda\,\partial_\tau-4\,\lambda^2\,x\,\partial_x+\left(y\,(-\lambda^2+1)+2\,\lambda\,z\right)\,\partial_y-\left(z\,(\lambda^2-1)+2\,\lambda\,y\right)\,\partial_z
\ee This metric was first given, although produced in a different
way, in \cite{Harrison}. Also in this case there are
special values of the constant $\lambda$ for which we have a fourth
Killing field: \be \lambda=\frac{\sqrt{3}}{3}\Rightarrow
\xi_4=6\,\partial_\tau+2\,\sqrt{3}\,x\,\partial_x-(\sqrt{3}\,y+3\,z)\,\partial_y+(3\,y-\sqrt{3}\,z)\,\partial_z\nonumber
\ee \be \lambda=-\frac{\sqrt{3}}{3}\Rightarrow
\xi_4=6\,\partial_\tau-2\,\sqrt{3}\,x\,\partial_x+(\sqrt{3}\,y-3\,z)\,\partial_y+(3\,y+\sqrt{3}\,z)\,\partial_z,\nonumber
\ee while there is \textbf{no} homothety. Finally, it is worth
noting that in  metric \eqref{Kasner II} the hypersurface
$t=\text{const.}$ is spacelike.

Another obtained member of the Harrison families is
\be\label{Kasner III}
d\,s^2=e^{(2\,\lambda+1)\,\tau}\,d\,\tau^2+e^{\tau}\,d\,x^2+\tau\,e^{\lambda\,\tau}\,d\,y^2+2\,e^{\lambda\,\tau}\,d\,y\,d\,z
\ee
This metric admits the homothetic field
\be
H=2\,\partial_\tau+2\,x\,\lambda\,\partial_x+(\lambda+1)\,y\,\partial_y+(z\,(\lambda+1)-y)\,\partial_z
\ee
In this case too, the hypersurface $t=\text{const.}$ is spacelike. Furthermore, for the value $\lambda=0$ the metric \eqref{Kasner III} describes a pp-wave, since the Killing field $u=\xi_3=\partial_z$ has zero covariant derivative and zero measure:
\be
\lambda=0\Rightarrow u^\alpha\,u_\alpha=0\,\wedge\, u^\alpha_{\ph{a};\beta}=0
\ee
\item \textbf{New metric}.
The metric given below is, to the best of our knowledge, new.
\be\label{Kasner IV}
d\,s^2=d\,t^2+2\,t^2\,d\,x^2+d\,y^2-4\,t\,d\,x\,d\,y+4\,d\,x\,d\,z
\ee
This metric admits the homothetic field $H=t\,\partial_t+y\,\partial_y+2\,z\,\partial_z$, describes a pp-wave since for the Killing field $u=\xi_3=\partial_z$, we have $u^\alpha\,u_\alpha=0\,\wedge\, u^\alpha_{\ph{a};\beta}=0$ and the hypersurface $t=\text{const.}$ is spacelike. It can also be proven that this metric is different from the pp-wave ensuing by setting $\lambda=0$ in \eqref{Kasner III}: indeed the tensor \be \Pi_{\alpha\beta\gamma\delta\epsilon}=
R^{\kappa\ph{a}\lambda\ph{b}}_{\ph{k}\alpha\ph{l}\beta}\,
R_{\kappa\gamma\delta\epsilon;\lambda}, \ee
vanishes identically for metric \eqref{Kasner IV} but not for metric
\eqref{Kasner III} with $\lambda=0$.
\end{itemize}
\subsection{Bianchi Type $II$}

In this model the structure constants, the basis 1-forms and the Killing fields are
\bsub
\begin{align}\label{structure constants II}
C^1_{\ph{1}23}=-C^1_{\ph{1}32}=\frac{1}{2} &\nonumber\\
C^\alpha_{\ph{a}\beta\gamma}=0&\quad \text{for all the other values of}\, \alpha, \, \beta, \, \gamma
\end{align}
\be\label{sigmaII}
\boldsymbol{\sigma}^1=z\,\boldsymbol{d}\,x+\boldsymbol{d}\,y,\quad
\boldsymbol{\sigma}^2=\boldsymbol{d}\,z,\quad
\boldsymbol{\sigma}^3=\boldsymbol{d}\,x
\ee
\be\label{killing type II} \xi_1=\partial_x,\quad
\xi_2=-x\,\partial_y+\partial_z,\quad \xi_3=\partial_y
\ee
\esub
\begin{itemize}
\item  \textbf{Lorenz-Petzold metric}.\be\label{Type II Euc}
d\,s^2=t\,\bs{d}\,t^2+\frac{1}{t}\,(\bs{\sigma}^1)^2+t\,(\bs{\sigma}^2)^2+t\,(\bs{\sigma}^3)^2
\ee
The above metric was first obtained in \cite{Lorenz 1983} and reproduced later by Valent \cite{Valent}. It is of \textbf{Euclidean} signature, admits a fourth Killing field $\xi_4=-2\,z\,\partial_x+(z^2-x^2)\,\partial_y+2\,x\,\partial_z$ and a homothety produced by the field $H=t\,\partial_t+x\,\partial_x+2\,y\,\partial_y+z\,\,\partial_z$.

Another obtained member of the Lorenz-Petzold families is
\be\label{Type II Euc non-hom}
d\,s^2=\mu^2\left(e^{-\xi}\,\xi\,\bs{d}\,\xi^2+\frac{1}{\xi}\,(\bs{\sigma}^1)^2+\xi\,(\bs{\sigma}^2)^2+e^{-\xi}\,\xi\,(\bs{\sigma}^3)^2\right).
\ee This metric does  \textbf{not} admit a homothety, so the
constant $\mu$ is essential and was first in \cite{Lorenz 1983} and reproduced later by Valent \cite{Valent}.

\item \textbf{Taub metrics}.
\be\label{Taub solution}
d\,s^2=\kappa^2\bigg(e^{2\,\xi\,\coth2\,\sigma}\,\cosh\xi\,\bs{d}\,\xi^2+\textrm{sech}\xi\,(\bs{\sigma}^1)^2+
e^{\xi\,\coth\sigma}\,\cosh\xi\,(\bs{\sigma}^2)^2 +\non\\e^{\xi\,\tanh\sigma}\,\cosh\xi\,(\bs{\sigma}^3)^2\bigg)
\ee
This solution was first produced in \cite{Taub}. It does not admit a homothety, hence the constant $\kappa$ cannot be absorbed. It is noteworthy, that in the limiting case $\sigma\rightarrow+\infty$, this metric possesses a fourth Killing field $\xi_4=-2\,z\,\partial_x+(z^2-x^2)\,\partial_y+2\,x\,\partial_z$.
\end{itemize}
\subsection{Bianchi Type $V$}

In this model the structure constants, the basis 1-forms and the Killing fields are
\bsub
\begin{equation}\label{structure constants V}
\begin{array}{ll}
C^1_{\ph{1}13}=-C^1_{\ph{1}31}=C^2_{\ph{2}23}=-C^2_{\ph{2}32}=\disfrac{1}{2}\\ C^\alpha_{\ph{a}\beta\gamma}=0 & \text{for all the other values of}\; \alpha, \, \beta, \, \gamma
\end{array}
\end{equation}
\begin{equation}
\bs{\sigma}^1=e^{-x}\,\bs{d}\,z,\,\bs{\sigma}^2=e^{-x}\,\bs{d}\,y,\,\bs{\sigma}^3=\bs{d}\,x
\end{equation}
\be\label{Killing fields V}
\xi_1=\partial_z,\quad \xi_2=\partial_y,\quad \xi_3=\partial_x+y\,\partial_y+z\,\partial_z
\ee
\esub
\begin{itemize}
\item \textbf{Joseph metrics}.
\begin{equation}\label{Joseph}
d\,s^2=\kappa^2\left(\frac{1}{4}\,\textrm{csch}^3\tau\,\bs{d}\,\tau^2+e^{-\sqrt{3}\,\tau}\,\,\textrm{csch}\tau(\bs{\sigma}^1)^2+\right.\\\left.e^{\sqrt{3}\,\tau}\,\,\textrm{csch}\tau(\bs{\sigma}^2)^2+ \textrm{csch}\tau(\bs{\sigma}^3)^2\right)
\end{equation}
This metric was first derived in \cite{Joseph} and does not admit a homothety, hence the constant $\kappa$ is essential.
\item \textbf{Flat space parameterization}.
\be\label{Flat Type V}
d\,s^2=-\bs{d}\,\tau^2+\tau^2\,(\bs{\sigma}^1)^2+\tau^2\,(\bs{\sigma}^2)^2+\tau^2\,(\bs{\sigma}^3)^2
\ee
This metric even though describes a flat space-time is "new" in the sense that it is a parametrization of it, as a Type V cosmological line
element. In \cite{Wainr}, at page 194 the Milne form of flat space-time is considered only as a Type $VII_h$ parametrization.
\item \textbf{New neutral signature metric}.
\be\label{neutral V}
d\,s^2=-\frac{1}{4\,t}\,d\,t^2+t\,d\,x^2-e^{-2\,x}(t-1)\,d\,y^2+2e^{-2\,x}\,d\,y\,d\,z+e^{-2\,x}(t+1)\,d\,z^2
\ee
This metric has neutral signature $(--++)$, admits a homothetic field $H=2\,t\,\partial_t+z\,\partial_y+y\,\partial_z$.
\item \textbf{New Euclidean signature metric}.
\begin{equation}\label{Eucl V}
d\,s^2=\lambda^2\left(\frac{1}{4}\,\textrm{sech}^3\xi\,\bs{d}\,\xi^2+e^{-\sqrt{3}\,\xi}\,\textrm{sech}\xi\,(\bs{\sigma}^1)^2+\right.\\
\left.e^{\sqrt{3}\,\xi}\,\textrm{sech}\xi\,(\bs{\sigma}^2)^2+\textrm{sech}\xi\,(\bs{\sigma}^3)^2\right)
\end{equation}
This metric does not admit a homothetic field, thus the constant $\lambda$ is essential.
\end{itemize}

\section{Discussion}
The present work completes the first phase of the programm initiated
in \cite{tchris3}, utilizing the automorphisms of the various
Bianchi Types as Lie-Point Symmetries of the corresponding
Einstein's Field Equations with the aim of uncovering their solution
space. The power of the method lies in the fact that it constitutes
a semi-algorithm which, if successfully applied, results in the
acquisition of the entire space of solutions. This successful
implementation had, so far, been carried out in the case of Bianchi
Type $III$ (\cite{ChrTer CQG}) and $VII_h$ (\cite{Ter Chr}; while
the present communication covers Types $I$, $II$, $IV$ and $V$. In
all cases considered where the general solution is expected to  have
three essential constants ($III$, $IV$ and $VII_h$), it is given in
terms of the sixth Painlev\'{e} transcendent $\mathbf{P_{VI}}$,
along the way with all the known particular solutions. It is
noteworthy that these known metrics are rediscovered in a systematic
way and without any extra assumption, in contrast to how they were
originally obtained. The case of Types $I$, $II$, and $V$  is
characteristic: The general solutions, not considered as such at the
time of their first derivation and containing 1 or 2 essential
constants, were produced with the aid of various simplifying
ansatzen in a time scale of half a century; Kasner (1921), Taub
(1951), Harrison (1959), Joseph (1966). Here, they are
comprehensively re-acquired along with the solutions not attributed
to any one else which, to the best of our knowledge, are
$\textbf{new}$: the Lorentzian Type $I$ pp-wave metric \eqref{Kasner IV}, the Type $V$ metrics \eqref{Flat Type V},
\eqref{Eucl V} (Euclidean signature)  and \eqref{neutral V} (Neutral signature). Metric \eqref{Flat Type V}
even though describes a flat space-time is "new" in the sense that
it is a parametrization of it, as a Type V cosmological line
element. In \cite{Wainr}, at page 194 the Milne form of flat
space-time is considered only as a Type $VII_h$ parametrization. The
production of metrics with Euclidean signature may, at first sight,
strike as odd; since our staring point is a line element of
Lorenzian signature. However, it is made possible by allowing the
lapse to be determined through the quadratic constrained equation
instead of prescribing it by an $\emph{ab initio}$ choice of time
gauge.

In the remaining Bianchi Types $VIII$, $IX$ and the exceptional $VI_h$ the number of existing automorphisms is not sufficient
to allow our method to reduce the problem to a single \textbf{second} order ODE in one unknown function, but rather to a \textbf{third} order one.
We strongly suspect it to be an equivalent form of the Chazy type equations. However the task of proving it involves the search for an appropriate   Lie-B\"{a}cklund transformation which is highly non-trivial and non-algorithmic. We plan to return if and when there is something concrete to be reported.

Some directions for future work include the application of the method in the presence of matter sources and/or in higher dimensions.

%%%%%%%%%%%%%%%%%%%%%%%%%%%%%%%%%%%%%%%%%%%%%%%%%%%%%%%%%%%%%%%%%%%%%%%%%%%%%%%%%%%%%%%%

\end{document}